\documentclass[prb,twocolumn]{revtex4}
\usepackage{amsmath}
\usepackage{amssymb}
\usepackage[dvips]{graphicx}

\begin{document}
\newcommand{\vc}{\mathbf}
\newcommand{\be}{\begin{equation}}
\newcommand{\ee}{\end{equation}}
\newcommand{\bk}{{{\bf{k}}}}
\newcommand{\br}{{{\bf{r}}}}
\newcommand{\bq}{{\bf{q}}}
\newcommand{\bQ}{{\bf{Q}}}
\newcommand{\bea}{\begin{eqnarray}}
\newcommand{\eea}{\end{eqnarray}}
\newcommand{\ra}{\rangle}
\newcommand{\la}{\langle}
\newcommand{\upa}{\uparrow}
\newcommand{\dna}{\downarrow}
\newcommand{\bS}{{\bf S}}
\newcommand{\E}{{\epsilon}}

\title{Self-consistent slave rotor mean field theory for strongly
correlated systems}
\author{E. Zhao}
\affiliation{Department of Physics, University of Toronto, Toronto,
Ontario M5S-1A7, Canada}
\author{A. Paramekanti}
\affiliation{Department of Physics, University of Toronto, Toronto,
Ontario M5S-1A7, Canada}
\begin{abstract}
%% changed: copying from the new arxiv abstract 
Building on work by Florens and Georges, we formulate and study a self-consistent slave rotor mean field theory for strongly correlated systems. This approach views the electron, in the strong correlation regime, as a composite of a neutral spinon and a charged rotor field. We solve the coupled spinon-rotor model self-consistently using a cluster mean field theory for the rotors and various ansatzes for the spinon ground state. We illustrate this approach with a number of examples relevant to ongoing experiments in strongly correlated electronic systems such as: (i) the phase diagram of the isotropic triangular lattice organic Mott insulators, (ii) quasiparticle excitations and tunneling asymmetry in the weakly doped cuprate superconductors, and (iii) the cyclotron mass of carriers in commensurate spin-density wave and U(1) staggered flux (or d-density wave) normal states of the underdoped cuprates. We compare the estimated cyclotron mass with results from recent quantum oscillation experiments on ortho-II YBCO by N. Doiron-Leyraud et al (Nature 447, 565 [2007]) which appear to find hole pockets in the magnetic field induced normal state. We comment on the relation of this normal ground state to Fermi arcs seen in photoemission experiments above Tc. This slave rotor mean field theory can be generalized to study inhomogeneous states and strongly interacting models relevant to ultracold atoms in optical lattices.
\end{abstract}
\maketitle

\section{Introduction}
It is well known that strong electronic correlations in systems such
as the transition metal oxides and ultracold atoms in optical lattices
can lead to a wealth of novel phases and
interesting phenomena not observed in conventional metals and
superconductors. A continuing theoretical challenge is to develop
tools to solve model Hamiltonians, such as the Hubbard model or the 
$t$-$J$ model, to gain a qualitative as well as quantitative 
understanding of the strong correlation regime of these models where
perturbative methods are no longer expected to work. A 
variety of different approaches have been developed to address this 
issue, and they can be classified roughly into two categories. The 
first category includes numerical methods such as
exact diagonalization which are good at capturing the short distance
physics \cite{dagotto-rmp}. A much more sophisticated approach 
is dynamical mean field
theory (DMFT) which maps the Hubbard Hamiltonian onto a quantum 
impurity model
self-consistently coupled to a dynamical bath \cite{dmft-rmp,cluster-rmp}. 
This approach
and its extensions, which include some degree of
spatial fluctuations, have led to significant
progress in our understanding of the Mott (metal-insulator)
transition as well as the short-range spin correlations and
superconductivity in doped Mott insulators \cite{bkyung,kancharla}. 
In the second category
are semi-analytical approaches including renormalized mean field
theory \cite{rmft-zhang,rmft-pwa,baskaran} 
and slave particle mean field theories. 
Slave-particle mean field theories, such as slave boson 
\cite{coleman,kotliar-liu,Lee-rmp} and slave 
rotor \cite{Florens} theories, are motivated 
by the observation that strongly interacting systems often display (i)
quite disparate charge and spin response with (ii) little evidence
of sharply defined electron-like quasiparticles. This observation
has led to the phenomenological spin-charge separation postulate
that electrons in many strongly correlated materials
may be better viewed as composites of `chargons' and
`spinons', which respectively describe the charge and spin degrees
of freedom. A physically transparent way to encapsulate this
idea is provided by the slave particle approaches which enlarge the 
Hilbert space of the physical electronic degrees of freedom in terms
of separate chargon and spinon Hilbert space. The unphysical 
states are then eliminated by enforcing constraints on the enlarged
Hilbert space. The strongly correlated model
then maps onto a model of interacting slave particles
coupled to gauge fields \cite{Lee-rmp,Z2}. 
This rewriting of the strongly
correlated electronic model is useful in situations where the
gauge fluctuations are weak. This assumption, that the gauge fluctuations
are weak and can be ignored, is a reasonable approximation for conducting 
(metallic or superconducting) states, broken symmetry insulating phases
such as the Neel antiferromagnet, or if the gauge fluctuations are
gapped \cite{Z2}. In other cases, we have to appeal to
comparisons with experiments to see if such slave particle mean field 
approaches provide a more useful starting point to think about the 
physics compared to perturbative analyses.

In this paper we present a further development in the slave particle
mean field approach. In conventional U(1) slave boson theory for
the ground state of the $t$-$J$ model \cite{kotliar-liu,Lee-rmp}, one is 
in the regime where the
on-site Coulomb repulsion $U\gg t$, and the bosons are assumed to be 
condensed with the boson fluctuations being ignored. However, the
boson model is a strongly interacting model which has to be solved
in order to obtain its ground state and excitations. Similarly, for
systems close to a Mott transition, number (charge) fluctuations are 
significant so it is more appropriate to describe the charge degree 
freedom using slave rotors \cite{Florens}. 
The rotor model is also nontrivial: its Hamiltonian 
resembles the Bose-Hubbard model and again it has to be solved. For
instance, at commensurate densities, the rotor model can undergo a Mott
transition leading to an electronic Mott transition in the
coupled spinon-rotor problem. We study a number of simple
situations and present results which suggest
that {\it solving} the rotor model and coupling it 
self-consistently to the spinons provides a reasonable mean field
framework to study strongly correlated electronic systems, and also
recovers U(1) slave boson mean field results for $U/t \to \infty$.

The rotor sector have been studied by several authors using single
site mean field theory as originally applied to the Bose-Hubbard
model \cite{fisher-mott,sheshadri}. We shall show that while the single 
site approximation
captures the essence of Mott transition \cite{Florens}, it is inadequate 
in some
respects and sometimes yields incorrect results. Going beyond this
requires the evaluation of short range spatial correlations. We 
propose to solve the rotor Hamiltonian by
diagonalizing it on a small cluster coupled to the rest of the
lattice self-consistently through an order parameter bath. 
We demonstrate that such a
cluster mean field theory can get rid of the aforementioned problems
of single site approximation while keeping the whole procedure
conceptually simple and computationally affordable. This constitutes
the main new aspect of the present work. This procedure is rather
general, and we expect it would also be useful for other strongly
interacting bosonic models, e.g., those which arise in the context 
of cold atoms in optical lattices. It can also be
adapted to address inhomogeneous electronic states. In this paper,
we apply this technique to address issues arising from some recent 
experiments in strongly correlated electronic systems.

The outline of this paper and a summary of the main results 
is as follows. We begin in sections II
with a review of the basic idea of the slave rotor approach and the
$tUJ$ model Hamiltonian. In section III we introduce the
self-consistent mean field theory of the rotor sector and spinon
sector. Sections IV and V discuss various mean field ansatzes for the
spinon Hamiltonian, and the cluster mean field theory for the
rotors. The remainder of the paper deals with applications of the 
coupled spinon-rotor mean field theory. In Section VI, we study a metal 
to Mott insulator transition as well as the possibility of a spin
liquid phase in a Hubbard-type model at half-filling on triangular
lattice. This model describes organic compounds, such as
$\kappa$-(ET)$_2$Cu$_2$(CN)$_3$, which feature a pressure-tuned Mott
transition \cite{shimizu,Kurosaki}. We show that in contrast to a single-site 
mean field theory,
the cluster approach leads to the possibility of a spin liquid state
between the metallic state and the antiferromagnetically ordered
insulator. Section VII illustrates the application of the mean field
theory to the superconducting state of doped cuprates. Specifically,
we find that properties of quasiparticle excitations such as their
spectral weight and Fermi velocity, which are measured in photoemission
experiments, are in qualitative agreement with variational
wavefunction and renormalized mean field theory results. We also
study the particle-hole asymmetry observed in recent tunneling
experiments \cite{davis-Rmap}
on the weakly doped cuprates, and make an estimate of
how this asymmetry varies with the cutoff energy scale in tunneling
asymmetry imaging. In section VIII, we study a strongly correlated
commensurate spin-density wave (SDW) state and a U(1) staggered flux or 
d-density wave (DDW) state
as possible underlying normal states of the underdoped cuprates. We
use the slave rotor mean field theory to estimate the effective
cyclotron mass of the carriers in these states. These results are
compared with recent high field quantum oscillation experiments on
ortho-II YBCO which shows evidence of hole pockets. We suggest that
doping dependent studies of the cyclotron mass should 
be able to distinguish between these two normal states.

\section{Model and slave rotor representation}

We consider the $tUJ$ model for electrons on a two dimensional
lattice \be H = - \sum_{i,j,\sigma} t_{ij} c^\dagger_{i\sigma}
c^{\vphantom\dagger}_{j\sigma} + U \sum_i n_{i\upa} n_{i\dna} + J
\sum_{\la i,j\ra} \bS_i\cdot\bS_j \label{hamiltonian} \ee Here
$t_{ij} \equiv t,t'$ represents hopping amplitudes to nearest and
next nearest neighbor sites respectively. $U>0$ is the local Coulomb
repulsion. $\bS_i \equiv c^\dagger_{i\alpha}
\vec{\sigma}_{\alpha\beta} c^{\vphantom\dagger}_{i\beta}$ is the
electron spin operator, where $\vec{\sigma} \equiv
(\sigma_x,\sigma_y,\sigma_z)$ are the Pauli matrices. $J>0$ is an
antiferromagnetic spin exchange between spins on neighboring sites.
Ordinarily, the superexchange $J$ derives from a combination of the
kinetic terms and the Hubbard repulsion in the regime $U \gg t,t'$,
with $J \sim 4 t^2/U$. Here, however, we retain an explicit $J$ term
in the model since we will analyze the spin physics of this model
below using mean field theory. The $J$ term is required in order to
reproduce the strong coupling antiferromagnetic spin correlations
within mean field theory (which decouples the spin and charge
degrees of freedom). Even for weak to moderate couplings, $U
\lesssim t$, the Hubbard repulsion leads to antiferromagnetic spin
fluctuations: in this regime, the $J$ term can be viewed as a way to
reproduce the weak coupling equal time spin structure factor within
mean field theory. The $tUJ$ model have been studied previously by
several authors \cite{marston-flux,fuchun-zhang,Ross-mackenzie}. 
We will solve this model using the slave rotor
method, for a range of densities around half-filling and for a range
of repulsive interactions $U$. For studying interaction driven
transitions such as the Mott transition, we will keep $J$ fixed for
simplicity. In order to study the doping dependent correlations at
fixed $U$, we set the antiferromagnetic exchange $J=4 t^2/U$.

We begin with a review of the slave-rotor description of strongly
correlated electron systems introduced by Florens and Georges 
\cite{Florens}. The
electron Hilbert space at a single lattice site has four states,
$|0\ra, |\upa\ra, |\dna\ra, |\upa\dna\ra$. That is, it could be
empty, or be occupied by a spin up electron, or a spin down
electron, or be doubly occupied. In the slave rotor representation,
the electron charge degree of freedom is described by a charged
rotor and the spin degree of freedom is described by a spin-1/2
spinon (fermion). Each of the four physical states then corresponds
to a direct product of the rotor state and the spinon state, \bea
|0\ra &\equiv& |1\ra|0\ra \nonumber \\
|\upa\ra&\equiv& |0\ra|\upa\ra \nonumber \\
|\dna\ra &\equiv& |0\ra|\dna\ra \nonumber \\
|\upa\dna\ra &\equiv& |-1\ra|\upa\dna\ra \nonumber \eea Here on the r.h.s.,
the first ket $|n^\theta\rangle$ is the eigenstate of rotor charge
with eigenvalue $n^\theta=0,\pm 1$, and the second ket is eigenstate
of spinon occupation number, $n_{f,\sigma}=0, 1$ for
$\sigma=\uparrow,\downarrow$. Notice we have chosen a background
charge $+1$ for the state with no electrons and each added electron
contributes charge $-1$. The enlarged rotor-spinon Hilbert space
contains unphysical states such as $|1\ra|\upa\ra$. These unphysical
states are avoided by imposing the operator constraint \be
n^{\theta}_i + n^f_{i,\upa} + n^f_{i,\dna} = 1. \ee In the slave
rotor representation, the electron number is equal to the spinon
number, i.e., \be n^{\rm e}_{i\sigma} = n^f_{i\sigma}. \ee The
electron creation (annihilation) operator \bea
c^\dagger_{i,\sigma}&=&f^\dagger_{i,\sigma} {\rm e}^{-i\theta_i} \label{cre},\\
c^{\vphantom\dagger}_{i,\sigma}&=&f^{\vphantom\dagger}_{i,\sigma}
{\rm e}^{+i\theta_i}, \label{ann} \eea where $f_{\sigma}$ is the
spinon annihilation operator, and the rotor creation (annihilation)
operator , ${\rm e}^{+i \theta_i}$ (${\rm e}^{-i\theta_i}$), is
defined by \bea {\rm e}^{\pm i\theta_i}
|n_i^{\theta}\ra&=&|n_i^{\theta} \pm 1\ra . \eea We can then rewrite
the $tUJ$ Hamiltonian in terms of the spinon and rotor field
operator as \bea H_{\rm SR} &=& - \sum_{i,j,\sigma} t_{ij}
f^\dagger_{i\sigma} f^{\vphantom\dagger}_{j\sigma} {\rm
e}^{-i\theta_i} {\rm e}^{+i\theta_j} \nonumber\\
&+& \frac{U}{2} \sum_i n^\theta_i (n^\theta_i - 1) + J \sum_{\la
i,j\ra} \bS^f_i\cdot\bS^f_j. \label{srham}
\eea Here we have expressed the Hubbard
repulsion between electron charges in terms of the rotors since only
the rotors carry charge. Similarly, the antiferromagnetic exchange
interaction is expressed in terms of the spinons, with $\bS^f_i
\equiv f^\dagger_{i\alpha} \vec{\sigma}_{\alpha\beta}
f^{\vphantom\dagger}_{i\beta}$ since only the spinons have a spin
quantum number. Finally, the electron density is determined via the
spinon density since $\la n^f \ra = \la n^{\rm e} \ra$ as mentioned
earlier. Notice that in the usual slave-boson formulation of the
$tJ$ model, the doubly occupied state is forbidden because $U/t 
\to \infty$. By comparison, double occupancy is allowed in the slave 
rotor theory, which makes it possible to investigate the effect of 
charge fluctuations for arbitrary values of $U/t$. The slave rotor
formulation was used by Florens and Georges to study the Mott
transition of the Hubbard model (using single site mean field theory
as well as a sigma model for the rotor fluctuations) and the Coulomb
blockade phenomenon in quantum dots \cite{Florens}. Lee and Lee 
have used the slave
rotor representation to study possible spin liquid phases near the
Mott transition in Hubbard model on the triangular lattice \cite{Lee-lee}.

\section{Slave Rotor Mean Field Theory}

In the spirit of slave particle mean field theory, we approximate
the electron wavefunction by the direct product of appropriate
spinon and rotor wavefunctions, with the constraint equation
satisfied on average in the resulting mean field state,$\la
n^\theta_i \ra + \la n^f_{i,\upa}\ra + \la n^f_{i,\dna} \ra = 1$.
Together with $\la n^{\rm e}\ra = \la n^f\ra$, this leads to two
mean field constraints, \bea
\sum_\sigma \la n^f_{i,\sigma}\ra &=& \la n^{\rm e} \ra \equiv 1-\delta, \label{cns1}\\
\la n^\theta_i \ra &=&  1 - \la n^{\rm e} \ra \equiv \delta,
\label{cns2}\eea where we have defined the doping $\delta$, with
$\delta>0$ ($\delta<0$) representing hole (electron) doping. For
example, the electron ground state $|\Psi\ra = |\Psi_\theta\ra
|\Psi_f\ra$, and our task reduces to find the normalized
wavefunctions $|\Psi_\theta\ra, |\Psi_f\ra$ subject to the
constraint that $\la \Psi_\theta| n^\theta_i |\Psi_\theta \ra =
\delta$ and $\la \Psi_f| n^f_i|\Psi_f\ra = 1-\delta$.

We define reduced spinon and rotor Hamiltonian as
\bea \tilde{H}_f &=& \la \Psi_\theta|  H_{\rm SR} |\Psi_\theta \ra \nonumber \\
&=& - \sum_{i,j,\sigma} t_{ij} f^\dagger_{i\sigma}
f^{\vphantom\dagger}_{j\sigma} B_{ij} +
\frac{U}{2} \sum_i \la n^\theta_i (n^\theta_i - 1) \ra_\theta \nonumber \\
&+& J \sum_{\la i,j\ra} \bS^f_i\cdot\bS^f_j \\
\tilde{H}_\theta&=&\la \Psi_f| H_{\rm SR} |\Psi_f \ra \nonumber \\
&=& - \sum_{i,j,\sigma} t_{ij} \chi_{ij} {\rm e}^{-i\theta_i} {\rm
e}^{+i\theta_j} + \frac{U}{2} \sum_i n^\theta_i
(n^\theta_i - 1)\nonumber \\
&+& J \sum_{\la i,j\ra} \la \bS^f_i\cdot\bS^f_j \ra_f. \eea Here we
have used the notation, $\la \ldots \ra_f \equiv \la
\Psi_f|\ldots|\Psi_f\ra$ and $\la \ldots \ra_\theta \equiv \la
\Psi_\theta|\ldots|\Psi_\theta\ra$, and also defined \bea
B_{ij} &=& \la {\rm e}^{-i\theta_i} {\rm e}^{+i\theta_j} \ra_\theta, \\
\chi_{ij}&=& \la f^\dagger_{i\sigma} f^{\vphantom\dagger}_{j\sigma}
\ra_f. \eea Note that there is no implicit summation over spin
$\sigma$ in defining $\chi_{ij}$, and we have assumed $\chi_{ij}$ is
spin-independent. We will denote the nearest and next-nearest
neighbor values of $B_{ij}$ as $B$ and $B'$ respectively. Similarly,
the nearest and next-nearest neighbor values of $\chi_{ij}$ will be
denoted by $\chi$ and $\chi'$.

The ground state energy in mean field theory is $E_0 = \la \Psi_f |
\tilde{H}_f | \Psi_f \ra = \la \Psi_\theta| \tilde{H}_\theta |
\Psi_\theta \ra$. Thus, in order to minimize $E_0$, we must choose
$|\Psi_f\ra$ to be the ground state of $\tilde{H}_f$, and
$|\Psi_\theta\ra$ to be the ground state of $\tilde{H}_\theta$. This
means we must self consistently solve for the ground state of the
two coupled Hamiltonians \begin{align} H_f &= -
\sum_{i,j,\sigma}\!\! t_{ij} B_{ij} f^\dagger_{i\sigma}
f^{\vphantom\dagger}_{j\sigma} + \!\!\! J \sum_{\la i,j\ra}
\bS^f_i\cdot\bS^f_j \!\! -\!\!
\!\mu_f \sum_{i\sigma} n^f_{i\sigma}, \\
H_\theta &=- 2 \sum_{i,j}\!\! t_{ij} \chi_{ij} {\rm
e}^{-i\theta_i} {\rm e}^{+i\theta_j}\!\! +\!\! \frac{U}{2} \sum_i
(n^\theta_i)^2 \!\!\!-\!\!\! \mu_\theta \sum_{i\sigma}\!\!
n^\theta_{i}, \end{align} where we have introduced chemical potentials for
the spinons ($\mu_f$) and rotors ($\mu_\theta$) as Lagrange
multipliers to impose the mean field number constraints
Eq.\eqref{cns1} and \eqref{cns2}. Note that the two decoupled
Hamiltonians $H_f,H_\theta$ are, at this stage, still strongly
interacting models, and we will use further approximations in the
spinon and rotor sectors to make progress. However, note that if we
develop better techniques to solve these independent spinon and
rotor Hamiltonians, we can obtain better results within this
self-consistent mean field approach.

The ground state energy of the electronic model is given by the
expectation value of the Hamiltonian $H_{\rm SR}$ of Eq.(\ref{srham}),
which leads to
\bea
E^0_{\rm e}&=& - \sum_{i,j,\sigma} t_{ij}
\chi_{ij} B_{ij}
+ \frac{U}{2} \sum_i \la n^\theta_i (n^\theta_i - 1)\ra_\theta 
\nonumber \\
&+& J \sum_{\la i,j\ra} \la \bS^f_i\cdot\bS^f_j \ra_f.
\eea

\section{Spinon Sector}
The spinon sector corresponds to a Hamiltonian with kinetic energy
and an antiferromagnetic exchange interaction. In principle, the
spinon sector can be treated using diagrammatic techniques,
especially when the spin interaction strength is small compared to
the spinon kinetic energy. Away from this regime, for larger spin 
interactions, the diagrammatic methods may not be adequate. Here we
adopt a simpler mean field approach to the spinon sector which has
the advantage that we can describe the normal state, various broken 
symmetry states, and singlet pairing states of spinons.

\subsection{Uniform spin liquid and $120^\circ$ Neel state on the
triangular lattice} There have been extensive theoretical studies of the Mott
transition on the triangular lattice 
\cite{capone,morita,parco,liu,gan,tremblay,aryan,Ross-mackenzie}. 
Among the suggested insulating
states are a uniform spin liquid which can be viewed as a `Fermi
liquid' of spinon quasiparticles \cite{Motrunich,Lee-lee}, 
and the $120^\circ$ Neel ordered
antiferromagnet. In order to study the Mott transition into such
states, we decouple the exchange interaction as \bea
\bS^f_i\cdot\bS^f_j  &\to&
 - \frac{3}{4} \chi_{ij}
\sum_\sigma \left(f^\dagger_{i\sigma} f^{\vphantom\dagger}_{j\sigma}
+ {\rm
h.c.}\right) \nonumber \\
&+& \frac{J M}{2} \left( {\rm e}^{i\bQ.\br_i} f^\dagger_{j\dna}
f^{\vphantom\dagger}_{j\upa} +
{\rm e}^{-i\bQ.\br_i} f^\dagger_{j\upa} f^{\vphantom\dagger}_{j\dna} \right.\nonumber \\
&+& \left. {\rm e}^{i\bQ.\br_j} f^\dagger_{i\dna}
f^{\vphantom\dagger}_{i\upa} + {\rm e}^{-i\bQ.\br_j}
f^\dagger_{i\upa} f^{\vphantom\dagger}_{i\dna}\right),
\label{sdwansatz} \eea where $\chi_{ij}$ is real and $M$ is the
sublattice magnetization which lies in the XY plane. The wavevector
$\bQ$ is chosen to correspond to the $120^\circ$ Neel state, namely
$\bQ\cdot\hat{a}=\bQ\cdot\hat{b}=2\pi/3$ where $\hat{a}\equiv
\hat{x}$ and $\hat{b}\equiv -\hat{x}/2+\hat{y}\sqrt{3}/2$ are unit
basis vectors on the triangular lattice. If $M=0$, the ansatz
reduces to a Fermi sea of spinons, while a nonzero $M$ leads to an
antiferromagnetically ordered state.

\subsection{D-wave pairing ansatz}
In order to describe the superconducting state of the cuprates, we
make a mean field ansatz for the spinon Hamiltonian, $H_f$,
following Kotliar and Liu \cite{kotliar-liu}. We decouple the antiferromagnetic
exchange term in the particle-particle and well as in the
particle-hole channel. Upto constant terms, this yields \bea
\bS^f_i\cdot\bS^f_j  &\to &
 - \frac{3}{4} \chi_{ij}
\sum_\sigma \left(f^\dagger_{i\sigma} f^{\vphantom\dagger}_{j\sigma}
+ {\rm
h.c.}\right) \nonumber \\
&-& \frac{3}{4} \Delta_{ij} \left( f^\dagger_{i\upa}
f^{\dagger}_{j\dna} - f^\dagger_{i\dna} f^{\dagger}_{j\upa} + {\rm
h.c.} \right). \label{ansatz-d-wave} \eea Here we assume $\chi_{ij}$
is real, and $|\Delta_{ij}| \equiv |\la
f^{\vphantom\dagger}_{i\dna}f^{\vphantom\dagger}_{j\upa}\ra|$ is
magnitude of the d-wave pairing, with $\Delta_{ij}$ being positive
on $x$-bonds and negative on $y$-bonds of the square lattice. This
decoupling scheme highlights the fact that there is no difference,
{\it at short distance}, between the tendency to antiferromagnetism
and the tendency to form d-wave pairs: both involve spin-singlet formation.
The difference between these states only becomes clear on longer
length scales, and will be dealt with in a subsequent paper.

\subsection{Commensurate spin-density wave ansatz}
As a candidate for the underlying normal state of the underdoped
cuprates, we examine a strongly correlated spin-density wave (SDW)
state with ordering wavevector $\bQ=(\pi,\pi)$, which is consistent
with that of the adjacent antiferromagnetic phase. To describe the
commensurate SDW ordered at $\bQ$, we decouple the exchange
interaction as \begin{align} \bS^f_i\cdot\bS^f_j  &\to
 - \frac{3}{4} \chi_{ij}
\left(f^\dagger_{i\sigma} f^{\vphantom\dagger}_{j\sigma} + {\rm
h.c.}\right)\nonumber  \\
&+ (-1)^{x_i+y_i} \frac{M}{2} \left(n^f_{j\upa} - n^f_{j\dna} -
n^f_{i\upa} + n^f_{i\dna}\right). \label{ansatz-sdw}\end{align} Here
$M(-M)$ refers to the expectation value of
$S_z=(n_{\upa}-n_{\dna})/2$ on the $A(B)$ sublattices.

\subsection{U(1) staggered flux or DDW ansatz}
Another candidate normal state for the weakly doped cuprates is the
so-called U(1) staggered flux \cite{marston-flux,ivanov-lee} or 
DDW state \cite{ddw-nayak,ddw}. This state
has orbital currents on the bonds of the square lattice which form a
staggered pattern and break time-reversal and lattice symmetries.
The mean field ansatz for this state is obtained by decoupling the
exchange interaction as \be \bS^f_i\cdot\bS^f_j  \to
 - \frac{3}{4}
\left(\chi_{ji} f^\dagger_{i\sigma} f^{\vphantom\dagger}_{j\sigma} +
{\rm h.c.}\right). \label{eq-flux-ansatz} \ee where
$\chi_{ji}=\chi_{ij}^*$ is complex, with a phase corresponding to an
enclosed flux of $0 \leq \Phi_f\leq \pi$ per plaquette. $\Phi_f$ is
staggered (alternating in sign) from one square plaquette to the
next in a checkerboard pattern. Further, we assume $\chi_{ij}$ is
real on diagonal bonds which connect sites of the same sublattice.

\section{Rotor Sector: Cluster mean field theory}

\begin{figure}
\includegraphics[width=3in]{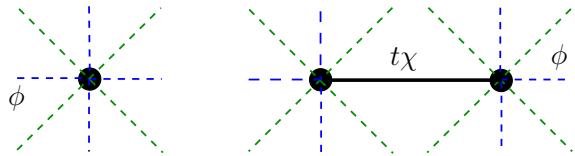}
\caption{Single site (left) and two-site cluster (right) for the
suqare lattice. Solid line shows nearest hopping within the
cluster. Dashed lines
represent coupling to the bath described by the order parameter
$\Phi$.} \label{fig-cluster}
\end{figure}

The rotor Hamiltonian can be solved by a variety of techniques used
commonly for bosonic systems. Here we formulate a self-consistent
cluster mean field theory which is a straightforward extension of
the single-site mean field theory used to describe the Mott
transition in the Bose-Hubbard model \cite{fisher-mott,sheshadri}. 
The idea of the cluster mean
field theory is to focus on a finite cluster of sites and treat the
influence of the sites outside the cluster (the ``bath") using a
mean field order parameter. The hopping terms that couple sites
within the cluster to sites outside the cluster are decoupled using
\be {\rm e}^{-i\theta_i}{\rm e}^{i\theta_j}\rightarrow {\rm
e}^{-i\theta_i}\Phi, \,\,\,\, i\in\mathrm{cluster},
j\in\mathrm{bath}, \ee where the mean field superfluid order
parameter $\Phi\equiv \la {\rm e}^{i\theta_i}\ra$ induces number
fluctuations in the cluster, and has to be determined
self-consistently. This is schematically shown in Fig.
\ref{fig-cluster} for the square lattice.

In the superfluid phase, $\Phi\neq 0$, and we choose it to be real.
The Mott insulator phase, on the other hand, is characterized by
$\Phi=0$ and a charge gap. All terms within the cluster, including
the onsite $U$ and intra-cluster hopping, are retained completely.
This cluster mean field Hamiltonian is diagonalized exactly to
obtain the eigen-energies and the ground state rotor wave function.
The explicit cluster Hamiltonian is given for specific cases below.
We note that by going to larger clusters, we can obtain a
progressively better description of the bosonic excitations of the
rotor sector.

%%%%%%%%%%%%%%%%%%%%%%%%%%%%%%%%%%%%%%%%%%%%%%%%%%%%%%%%%%%%%%%%%%%%%
%%%%%%%%%%%%%%%%%%%%%%%%%%%%%%%%%%%%%%%%%%%%%%%%%%%%%%%%%%%%%%%%%%%%%

\noindent{\bf Single-site mean field theory}: Applying the mean
field theory to a single-site ``cluster",  $H_\theta$ for a square
lattice model is reduced to a single site rotor Hamiltonian, \be
H^{\Box,(1)}_\theta = - 8 (t \chi\!\! +\!\! t' \chi') \Phi ({\rm
e}^{-i\theta} \!\!+\!\! {\rm e}^{+i\theta})\!\! +\!\! \frac{U}{2}\!
(n^\theta)^2\!\! - \!\!\mu_\theta n^\theta. \ee For given $\chi,
\chi'$ (obtained from the spinon problem) and charge density
$\langle n^\theta\rangle=\delta$, we start from some trial value of
$\Phi$ and diagonalize $H^{(1)}_\theta$ in the rotor number (angular
momentum) basis $\{|n^\theta\rangle \}$ by truncating the rotor
Hilbert space to $|n^\theta|\leq L$. $L$ is increased until the
results converge. We found $L=2$ is accurate enough for all of our
calculations. Using the ground state wavefunction $|G\ra_{(1)}$, the
ground state average $\langle n^\theta\rangle_{(1)}$ and $\Phi=\la
{\rm e}^{i\theta} \ra_{(1)}$ are computed. $\mu_\theta$ is adjusted
to give the correct charge density, and the procedure is iterated
until $\Phi$ is converged. Alternatively, one may minimize the
ground state energy under the charge density constraint $\langle
n^\theta\rangle_{(1)}$ to find $\Phi$, and the result will be the
same. Once $\Phi$ is determined self-consistently, we record energy
eigenvalues $E_\alpha$ and eigen states $|\alpha\ra$. Within the
single site approximation, the rotor kinetic energy $B = B' =
\Phi^2$. These values will be fed into the spinon Hamiltonian as
parameters. For the triangular lattice model with $t'=0$, the
corresponding single-site Hamiltonian is \be
H^{\triangle,(1)}_\theta = - 12 t \chi \Phi ({\rm e}^{-i\theta} +
{\rm e}^{+i\theta}) + \frac{U}{2} (n^\theta)^2 - \mu_\theta n^\theta.
\ee

\noindent{\bf Two-site cluster}: Going to larger cluster size yields
much better approximations for the rotor kinetic energy and
intersite correlations. For example, in a two site cluster model of
the the square lattice problem, the Hamiltonian $H_\theta$ is
approximated by \bea H^{\Box,(2)}_\theta&=& - 2 t \chi ({\rm
e}^{-i\theta_1} {\rm e}^{+i\theta_2}  + {\rm h.c.}) \nonumber \\
&-& (6 t \chi + 8 t' \chi_2)
\Phi ({\rm e}^{-i\theta_1} + {\rm e}^{-i\theta_2}  + {\rm h.c.})\nonumber  \\
&+& \frac{U}{2} (n^\theta_1)^2 + \frac{U}{2} (n^\theta_2)^2 -
\mu_\theta (n^\theta_1 + n^\theta_2). \eea Similar to the single
site case, this Hamiltonian is diagonalized numerically in basis
$\{|n^\theta_1,n^\theta_2\ra\}$ and $\Phi$ is determined
self-consistently via $\Phi = |\la {\rm e}^{\pm i\theta_1}\ra_{(2)}|
= |\la{\rm e}^{\pm i\theta_2}\ra_{(2)}|$, where the average is with
respect to the ground state $|G\ra_{(2)}$. Intersite correlations
are obtained from the ground state wavefunction using $B \equiv \la
{\rm e}^{-i\theta_1}{\rm e}^{+i\theta_2} \ra_{(2)}$ for nearest
neighbors and $B' \equiv \Phi^2$ for next nearest neighbors. These
will be used in the spinon Hamiltonian. For the triangular lattice,
the two site cluster Hamiltonian is given by \bea
H^{\triangle,(2)}_\theta&=& - 2 t \chi ({\rm
e}^{-i\theta_1} {\rm e}^{+i\theta_2}  + {\rm h.c.}) \nonumber \\
&-& 10 t \chi
\Phi ({\rm e}^{-i\theta_1} + {\rm e}^{-i\theta_2}  + {\rm h.c.})\nonumber  \\
&+& \frac{U}{2} (n^\theta_1)^2 + \frac{U}{2} (n^\theta_2)^2 -
\mu_\theta (n^\theta_1 + n^\theta_2). \eea Finally, we note that the
rotor hopping $t\chi_{ij}$ is complex in the U(1) staggered flux state.
The rotor Hamiltonian in this case can be solved similarly on a
two-site lattice to obtain the complex $B_{ij}$. It is straightforward
to generalize this to larger clusters depending on the available
computational resources.

The main qualitative difference in the results obtained from the
single-site mean field theory and the cluster mean field theories is
the following. In single site mean field theory, the kinetic energy
$B$ is the same as the square of the order parameter, $\Phi^2$.
Thus, it vanishes in the Mott insulating phase. This amounts to
neglecting all number fluctuations within the Mott phase, and it is
too crude an approximation especially when close to the Mott
transition. By contrast, the cluster mean field theory does much
better job capturing the short distance correlations, the kinetic
energy $\propto B$ does not simply follow $\Phi^2$. For instance at
the commensurate filling $\la n^\theta_i \ra = 1$ and large $U/t$,
deep in the Mott insulator, we find $B \sim t/U$.

\section{Mott transition on the triangular lattice at half-filling}
A Mott transition refers to an interaction induced transition from
a conducting phase to an insulating phase. The insulating phases
may or may not be accompanied by any conventional broken
symmetries. 
One interesting recent example of such a Mott transition
is the pressure induced insulator to metal
transition in organic compounds such as $\kappa$-(ET)$_2$Cu$_2$(CN)$_3$
at an effective density of one electron per site (half-filling) 
\cite{shimizu,Kurosaki}.
The simplest model Hamiltonian describing this material is the Hubbard
model on an isotropic triangular lattice. Here we use the cluster
slave rotor mean field theory to study the $T=0$ phase diagram of
model Eq. \eqref{hamiltonian} on a triangular lattice with $t'=0$.

Previous exact diagonalization studies of the triangular lattice Hubbard 
model suggested \cite{capone} a direct first order transition from a 
uniform metallic phase
into an antiferromagnetic insulator with $120^\circ$ Neel order at
$U/t \approx 12$. However that study also pointed out \cite{capone}
that a more
complex phase diagram emerges from Hartree-Fock calculations which
show a metallic SDW and incommensurate spiral ordering in addition to
the non-magnetic metal and the antiferromagnetic insulator. However,
the validity of such a weak-coupling approach is unclear for such
large values of $U/t$.
More recent numerical studies \cite{morita}
using the so-called
``path integral renormalization group" (PIRG) method indicate that the
ground state first undergoes a metal to spin-liquid insulator transition 
at a critical $U_c \sim 5t$. At larger $U > U_c$, the spin liquid 
phase undergoes a second transition into the $120^\circ$ Neel ordered 
antiferromagnetic
state. Variational wave function studies, which attack the problem from
the insulating phase, find that the $120^\circ$ Neel ordered insulator
is unstable to a spin liquid state which may resemble a projected Fermi sea 
of spinons \cite{Motrunich}. This spin liquid state with a Fermi sea of 
spinons has been argued \cite{Motrunich,Lee-lee}
to provide a description of the non-magnetic insulating phase
seen in $\kappa$-(ET)$_2$Cu$_2$(CN)$_3$.

Using the ansatz Eq. \eqref{sdwansatz} for the triangular lattice,
we can obtain the mean field spinon Hamiltonian \begin{align}
H^{\mathrm{fs}}_f &= \sum_{\bk,\sigma} \xi_\bk
f^\dagger_{\bk\sigma} f^{\vphantom\dagger}_{\bk\sigma}
- \frac{3 J M}{2} \sum_\bk \left( f^\dagger_{\bk\upa}
f^{\vphantom\dagger}_{\bk-\bQ\dna} +\mathrm{h.c.} \right), \\
\xi_\bk &= -2(tB + \frac{3 J \chi}{4})(\cos k_a+\cos k_b+\cos k_c) -
\mu_f .
\end{align} where
$k_a=\bk\cdot\hat{a}$, $k_b=\bk\cdot\hat{b}$, and $k_c=k_a+k_b$. The
spinon kinetic energy has contributions from the bare spinon hopping
as well as from the antiferromagnetic interactions.

The excitation energy of the SDW quasiparticles is
given by \bea \lambda^{\pm}_\bk=\frac{\xi_\bk+\xi_{\bk-\bQ}}{2} \pm
\frac{1}{2} E_\bk ,\\
E_\bk\equiv\sqrt{(\xi_\bk-\xi_{\bk-\bQ})^2 + (3 J M)^2}. \eea
The self-consistency conditions for $M,\chi$ and the number
constraint equation $\la n_f \ra = 1-\delta$ (which determines
$\mu_f$) reduce to \bea M&=&\frac{1}{N} \sum_\bk \frac{3 J M}{2
E_\bk} \left[\Theta(-\lambda^-_\bk)
- \Theta(-\lambda^+_\bk)\right] ,\\
\chi&=&\frac{1}{3 N} \sum_\bk \left[u^2_\bk \Theta(-\lambda^-_\bk)
+ v^2_\bk \Theta(-\lambda^+_\bk) \right] \nonumber \\
&\times& (\cos k_a + \cos k_b + \cos k_c),  \\
1-\delta&=&\frac{1}{N} \sum_\bk \left[\Theta(-\lambda^+_\bk) +
\Theta(-\lambda^-_\bk) \right]. \eea Here $\Theta(x)$ is the step
function, and the coherence factor
\be u^2_\bk = \frac{1}{2} \left[ 1 -
\frac{\xi_\bk-\xi_{\bk-\bQ}}{E_\bk} \right],\;\;\;
v^2_\bk = 1-u^2_\bk. \ee

%%%%%%%%%%%%%%%%%%%%%%%%%%%%%%%%%%%%%%%%%%%%%%%%%%%%%%%%%%%%%%%%%%%%%
\begin{figure}
\includegraphics[width=3in]{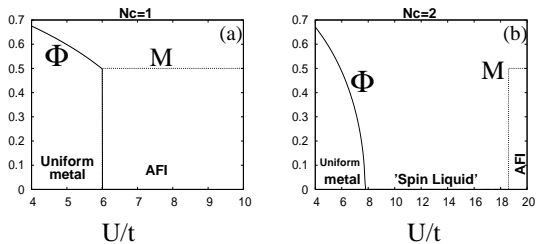}
\caption{The rotor order parameter $\Phi$ and the sublattice magnetization $M$
as function of $U/t$ for the triangular $tUJ$ model at half filling, $J=t/4$.
Left (a): Predictions of the single-site slave rotor mean field theory. The
transition from the uniform metal to the antiferromagnetic insulator (AFI)
is first order. Right (b): Results obtained using a 2-site rotor cluster. There
is an intermediate spin liquid state in which $M$ is zero within our
numerical accuracy.
}\label{fig-tr-mott}
\end{figure}
%%%%%%%%%%%%%%%%%%%%%%%%%%%%%%%%%%%%%%%%%%%%%%%%%%%%%%%%%%%%%%%%%%%%%

The results obtained for the Mott transition with a fixed $J = t/4$
are plotted in Fig.\ref{fig-tr-mott}. We show results for the rotor
order parameter $\Phi$, which distinguishes metallic ($\Phi\neq
0$) from insulating ($\Phi=0$) states, and the sublattice
magnetization $M$ for the $120^\circ$ Neel state. Using a single-site
cluster for the rotors, and computing the energy of the different states,
we find the phase diagram in Fig.\ref{fig-tr-mott}(a) which shows
a direct, strongly
discontinuous, transition from the uniform metal into a magnetically 
ordered insulating state, occurring at $U/t \sim 6$. The magnetic 
ordering in the insulator is
`classical' in the sense that the ordered moment is the full moment
($M=1/2$), 
and the electronic kinetic energy vanishes in the insulator. 
By contrast, the 2-site 
cluster for the rotor yields a richer phase diagram as shown in
Fig.\ref{fig-tr-mott}(b). It exhibits a
continuous Mott transition into an insulating state at about
$U/t \sim 7.8$. However the insulating state has no ordered moment,
$M$ is essentially zero within our numerical accuracy.
The insulating phase is thus consistent with a 
featureless `spin liquid' described by a spinon Fermi sea. This
`spin liquid' state may be further unstable to spinon pairing but we
have not, thus far, explored such a possibility.
For larger $U/t \sim 18.5$, we find a
discontinuous transition of this uniform `spin liquid' state into a 
$120^\circ$ ordered antiferromagnet. This result, of a transition 
at large $U/t$ from the spin liquid into the ordered state, can be checked 
using PIRG calculations extended to larger values
of $U/t$ than in Ref.~\onlinecite{morita}. Our main message is that
using a cluster approach to the rotor sector leads to a significant
qualitative difference with respect to a single-site mean field theory
in that it leads to an insulating spin liquid phase at intermediate
$U/t$.

This phase diagram is not too sensitive to the precise value of
$J/t$ so long as it is not too large. For smaller values
of $J/t$, the window of the `spin liquid' phase extends to larger
$U/t$. For much larger values of $J/t$, we find that even the larger
clusters can yield a strong discontinuous transition from the metal
to a magnetically ordered insulator. The details of the phase
diagram in the slave rotor mean field theory are determined by the
competition between the renormalized spinon kinetic energy, $\sim
t^2/U$, and the {\it fixed} exchange constant $J$. If the Mott
transition of the rotors occurs at a small $U/t$, then the
renormalized spinon kinetic energy is large compared to the
antiferromagnetic exchange constant $J$ and this can frustrate
magnetic ordering. This is similar in spirit, although not in details,
to local charge fluctuations frustrating
magnetic order on the triangular lattice via ring-exchange processes. 
Using a variational approach to the ring-exchange model, Motrunich has 
shown that it can significantly suppress or completely destroy magnetic 
order leading to a uniform spin liquid state \cite{deficiency}. A phase
diagram for the triangular lattice Mott insulator with an
intervening spin liquid phase is also consistent with results from
numerical approaches such as the PIRG. However the metal-insulator
transition point obtained in that study is at $U/t \sim 5$
indicating that our mean field result may not be quantitatively accurate.

\section{d-wave superconductivity in the doped cuprate materials}

High temperature superconductivity in the hole doped cuprates
has been a prototypical example of how strong correlations lead
to a variety of novel phenomena not observed in conventional superconductors.
There are two, seemingly quite different, approaches to d-wave
superconductivity in the cuprates. One approach starts from an
electronic Fermi liquid and shows that antiferromagnetic spin fluctuations 
centered at momentum $(\pi,\pi)$ on a square lattice
can mediate d-wave pairing of electrons
\cite{spin-fluct,scalapino,miyake}. The
other point of view, proposed initially
by Anderson, \cite{pwa-87} begins from the 
undoped insulating state of the
cuprates and argues that this is close to forming a spin liquid with
d-wave singlet pairing which naturally leads to d-wave superconductivity
upon doping away from half-filling.

From the perspective of the coupled spinon-rotor theory these two
approaches may be reconciled at low energy.
The weak coupling spin-fluctuation
calculations would, when applied to the spinon sector of the
slave rotor mean field theory,
predict an instability of {\it spinons} towards d-wave pairing 
arising from antiferromagnetic fluctuations induced by $J$.
Thus, although conventional spin fluctuation theories and models such
as the spin-fermion model \cite{chubukov} are
quite different from spin liquid based approaches, they appear to
be formally identical when reinterpreted as applying to spinons
rather than electrons.
At low dopings, when the exchange interaction becomes much stronger 
than the spinon kinetic energy, the weak coupling `spinon pair' may 
smoothly cross over into the resonating valence bond singlets 
of the spin liquid Mott state. Naively, this crossover is expected 
to happen 
when the kinetic energy of the doped carriers becomes comparable to the
exchange energy between neighboring spins.
Recent cellular DMFT calculations \cite{haule}
suggest that this crossover may
exhibit some features of `local quantum criticality'.

There has been extensive work
on the $tJ$ model using U(1) and SU(2) slave boson mean field 
(and corresponding gauge) theories \cite{baskaran,kotliar-liu,Lee-rmp}.
The slave boson mean field theory successfully captures some
essential features of the high temperature superconductors, 
including the d-wave symmetry, the pseudogap energy scale, and the 
superconducting dome \cite{kotliar-liu}. Despite the seemingly bold
approximations involved in such a slave particle formulation,
the qualitative predictions of the slave boson mean field theory 
agree remarkably well with more sophisticated methods like the variational 
wave function approach \cite{arun-prl,arun-prb}. 

In this section, we apply
the cluster slave rotor mean field theory to study the superconducting 
phase of the hole-doped square lattice $tUJ$ model. 
Our first purpose is to check that it can qualitatively
reproduce the well known
results of the $tJ$ model in the regime $U\gg t,J$. Second, 
we show that by {\it solving} the rotor problem, we learn about the 
incoherent spectrum weight associated with charge fluctuations which 
leads to, among other things, particle-hole tunneling asymmetry.

The d-wave pairing ansatz for the spinon sector leads to the well
known Hartree-Fock-Bogoliubov Hamiltonian \bea H^{\mathrm{dsc}}_f
&=& \sum_{\bk,\sigma} \xi_\bk f^\dagger_{\bk\sigma}
f^{\vphantom\dagger}_{\bk\sigma} - \sum_{\bk} (\Delta_\bk
f^\dagger_{\bk\upa} f^{\dagger}_{-\bk\dna} +
\mathrm {h.c.}) ,\\
\xi_\bk &=& -2 (t B + 3 J \chi/4)(\cos k_x + \cos k_y) \nonumber \\
&-& 4  t'B' \cos k_x \cos k_y - \mu_f, \\
\Delta_\bk &=& \Delta_0 (\cos k_x - \cos k_y)/2. \eea $H_f^{\rm
dsc}$ is diagonalized by the standard Bogoliubov transformation to
yield quasiparticle excitation energies
$E_\bk=\sqrt{\xi^2_\bk+\Delta^2_\bk}$. Here $\Delta_0=3J
|\Delta_{ij}|$ is the antinodal gap. $\Delta_\bk$, $\chi$, and
$\mu_f$ are determined self-consistently by \bea
1-\delta &=& \frac{1}{N}\sum_\bk(1-\frac{\xi_\bk}{E_\bk}),\\
1&=&\frac{1}{N}\sum_k\frac{3J}{8E_\bk}(\cos k_x-\cos k_y)^2,\\
\chi&=&\frac{1}{4N}\sum_k(1-\frac{\xi_\bk}{E_\bk}) (\cos k_x+\cos
k_y). \eea After self-consistency is achieved, \be
\chi'=(2N)^{-1}\sum_k(1-{\xi_\bk} /{E_\bk})\cos k_x \cos k_y \ee is
computed as input to the rotor sector. 

%%%%%%%%%%%%%%%%%%%%%%%%%%%%%%%%%%%%%%%%%%%%%%%%%%%%%%%%%%%%%%%%%%%%%
\begin{figure}
\includegraphics[width=1.6in]{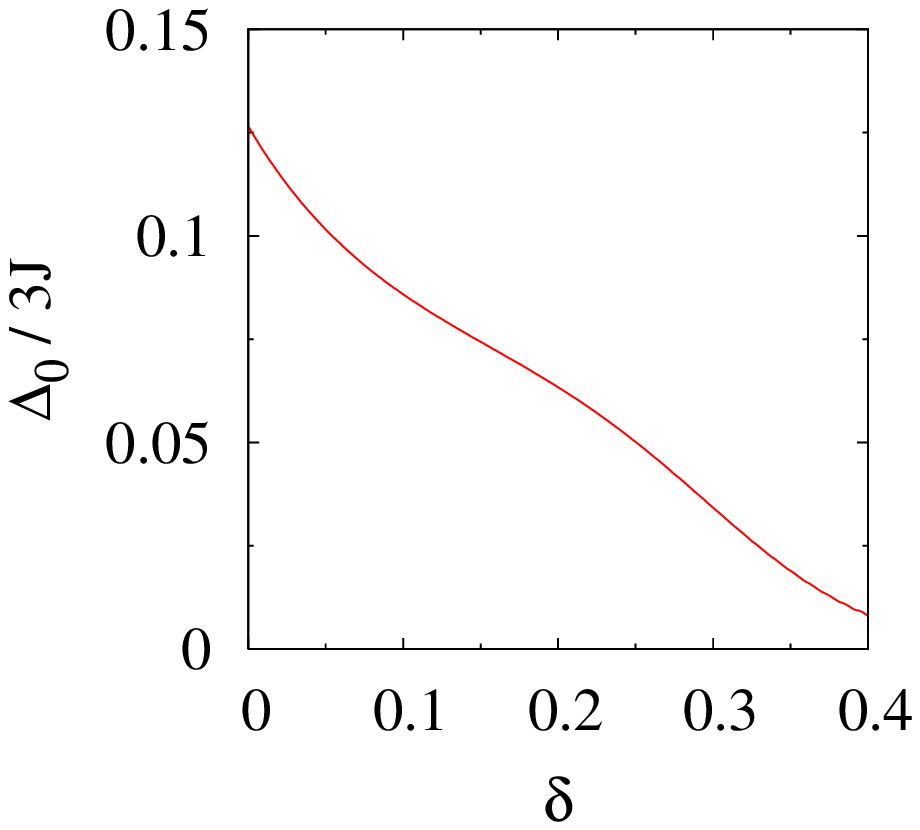}
\includegraphics[width=1.6in]{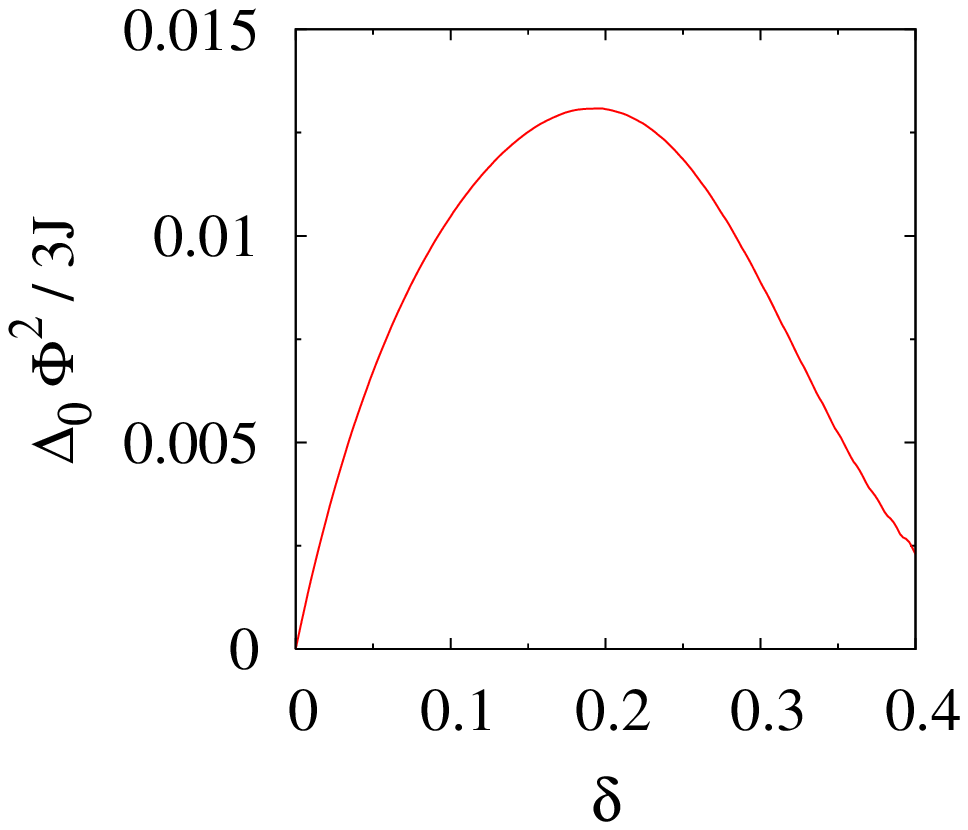}
\caption{Doping dependence of d-wave superconductivity arising
from doping a Mott insulator. Left: Maximum superconducting
gap $\Delta_0$. Right: Superconducting order parameter
$\langle c_{i\uparrow}c_{j,\downarrow}\rangle$.}
\label{fig-gap-op}
\end{figure}
%%%%%%%%%%%%%%%%%%%%%%%%%%%%%%%%%%%%%%%%%%%%%%%%%%%%%%%%%%%%%%%%%%%%%

We have numerically solved
the coupled rotor-spinon problem for $U=12t$, $J=t/3$, $t'=-t/4$
(these values are typical for hole doped cuprate high temperature
superconductors) at doping $0 \leq \delta\leq 0.4$ using a two site
rotor cluster. 
Fig. \ref{fig-gap-op} shows the doping dependence of the
antinodal spinon gap $\Delta_0$ as well as the superconducting order
parameter $\langle c_{i\uparrow}c_{j,\downarrow}
\rangle=\Phi^2\Delta_0/3J$. The spinon gap decreases monotonically
with increasing doping and corresponds to the psuedogap observed in
high temperature superconductors. The doping dependence of the order
parameter, on the other hand, takes a dome shape, reaching maximum
at optimal doping $\delta\sim 0.2$. This is in accordance with the
doping evolution of $T_c$ of high temperature superconductors. The
demise of superconductivity at low doping is due to the vanishing of
condensate density, which in the slave rotor formulation is
proportional to $\Phi^2$ and goes roughly as $\delta$ at low
dopings.

\subsection{Doping dependence of quasiparticle weight and nodal Fermi
velocity}

%%%%%%%%%%%%%%%%%%%%%%%%%%%%%%%%%%%%%%%%%%%%%%%%%%%%%%%%%%%%%%%%%%%%%
\begin{figure}
\includegraphics[width=1.6in]{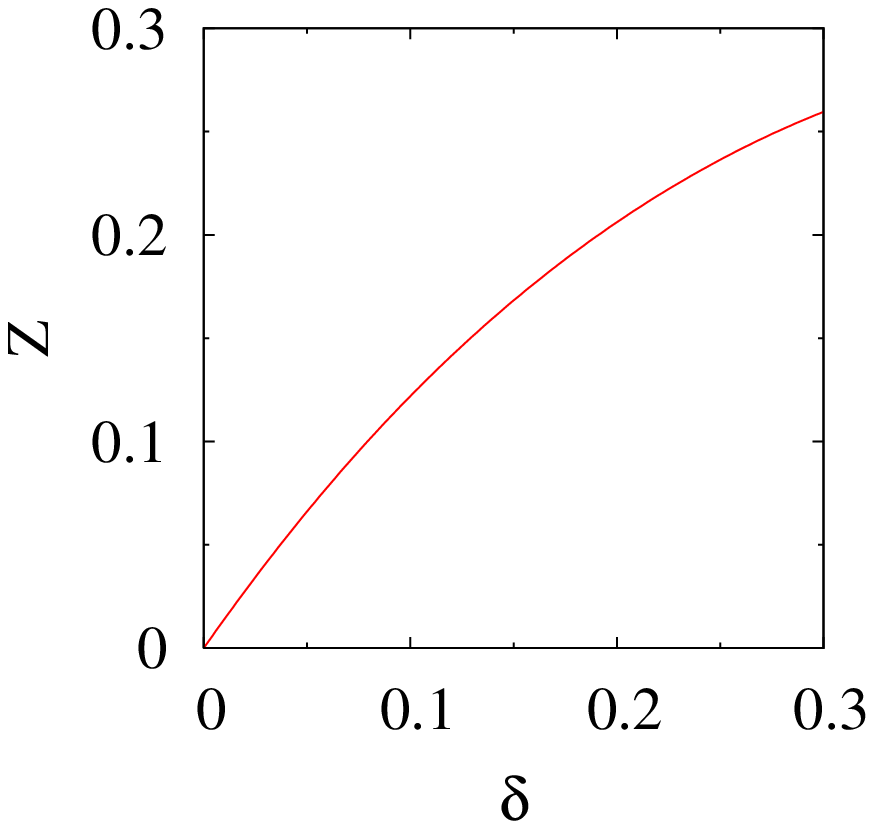}
\includegraphics[width=1.6in]{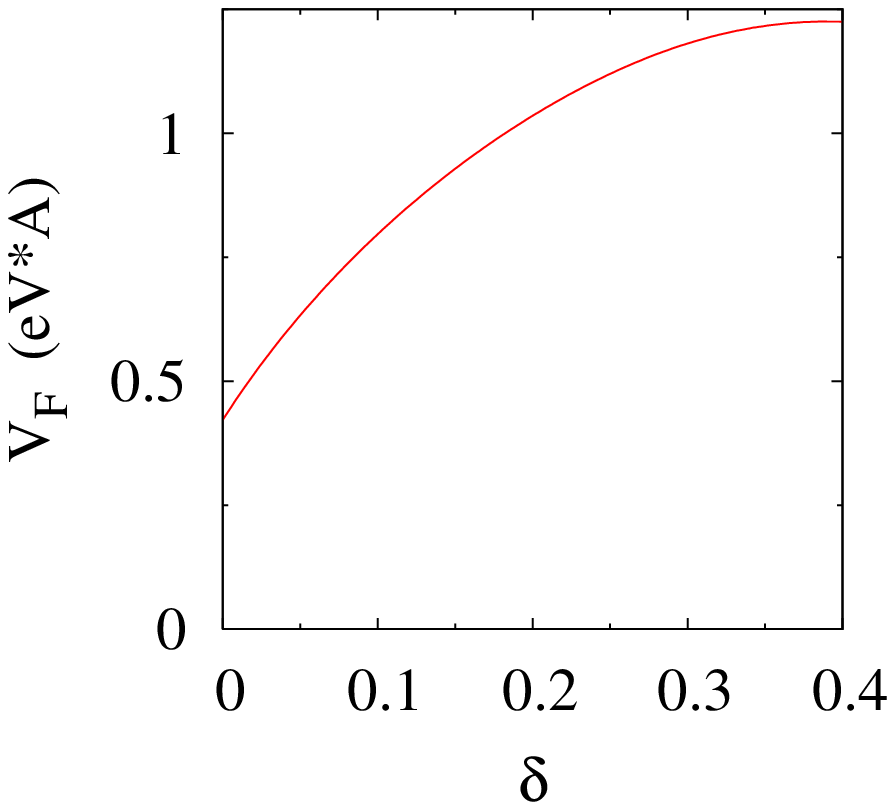}
\caption{Quasiparticle properties of the d-wave superconductor. 
Left: the quasiparticle weight $Z$ obtained from
slave-rotor mean field theory. Right: nodal Fermi velocity $V_F$ as 
function of doping.}\label{fig-z-vf}
\end{figure}
%%%%%%%%%%%%%%%%%%%%%%%%%%%%%%%%%%%%%%%%%%%%%%%%%%%%%%%%%%%%%%%%%%%%%

Interaction effects result in a transfer of spectral weight from 
sharply defined coherent quasiparticles to incoherent excitations. This
reduces the quasiparticle weight $Z$ from unity. Within the slave
rotor mean field theory, $Z$ is given by $\Phi^2$ and plotted in the
left panel of Fig. \ref{fig-z-vf}. 
Our slave rotor result is close to the slave boson result, as expected 
since $U/t \gg 1$ and the number (charge) fluctuation is suppressed.
In the right panel of Fig. \ref{fig-z-vf} we plot the nodal
quasiparticle Fermi velocity, $v_F=\partial \xi_k/\partial k$ for
$k=k_F$ along the nodal direction. Here in order to measure $v_F$ in
unit of eV-$\mathrm{\AA}$ we have used $t=350$meV and a lattice
spacing 3.8$\mathrm{\AA}$. We find that $v_F$ is a smooth function
of $\delta$ and does not vanish as $\delta \to 0$. The doping
dependence of $Z$ and $v_F$ is in qualitative agreement with 
experiments.

However, slave boson and slave rotor theories underestimate $Z, v_F$ 
when compared with Gutzwiller 
projected variational wavefunction calculations \cite{arun-prl,arun-prb}, 
indicating that gauge fluctuations that bind slave particles have to be taken 
into account to describe those calculations. Furthermore, the
calculated value of $v_F$ is smaller (by a factor of two) compared with 
experiments
\cite{kaminski,lanzara}, while the doping dependence of $v_F$ in the 
range $\delta \sim 
0.1-0.2$ disagrees with the experimental observation \cite{lanzara}
that $v_F\simeq 1.5$eV-$\mathrm{\AA}$ is essentially completely doping 
independent in this range.  
On the other hand good agreement with experiments is achieved 
by variational wave function calculations \cite{arun-prl,arun-prb}. These
disagreements show a shared weakness of the slave-particle 
mean field theories. Namely some degree of spinon-chargon interaction 
beyond mean field theory, arising from gauge fluctuations, needs to be 
retained in order to make a {\it quantitative} comparison with 
the photoemission experiments.

\subsection{Particle-hole asymmetry in tunneling}
Tunneling experiments in the underdoped cuprate superconductors
find a marked asymmetry between the tunneling conductance at
positive bias (tunneling in an electron from tip to sample) versus
negative bias (extracting an electron from the sample). This had been
observed in earlier experiments in slightly underdoped materials 
\cite{Renner,pan} which focussed on the tunneling conductance at
low energy. More recent experiments, which have gone to lower doping
and to larger bias, have shown that this asymmetry grows significantly
in the strongly underdoped regime \cite{davis-Rmap}. 

This asymmetry was explained by calculations using slave boson mean field 
theory, which pointed out that it arises from the proximity of the
underdoped cuprates to a correlated insulating state at zero doping 
\cite{rantner}.  Rantner and Wen \cite{rantner, wen-asymmetry}
calculated the expected tunneling spectrum 
assuming that the slave bosons behave as free particles with a bandwidth 
$\sim t$. Here we use the self-consistent
slave rotor mean field theory to calculate the tunneling spectrum into
the weakly doped superconducting state of the cuprates. We show that
the tunneling spectrum has a coherent BCS-like form at low energies
while it has significant incoherent weight at negative bias in agreement
with earlier results. We also compute
the ratio of the integrated (upto a cutoff)
tunneling density of states at positive to 
that at negative bias and
compare the doping and cutoff dependence of this ratio with
recent tunneling experiments \cite{davis-Rmap}. 

Within the slave rotor mean field approach, using Eq. \eqref{cre}
and \eqref{ann}, the electron Green function
factorizes into spinon and rotor Green functions. This leads to
a simple qualitative picture for the electronic
tunneling process --- namely, tunneling an electron involves tunneling
a spinon and a charge. This factorization
preserves the sum rule on
the electron spectral function $A(\bk,\omega)$ that 
$\int_{-\infty}^{+\infty} d\omega A(\bk,\omega) = 1$. The
resulting {\it local} electron Green function \begin{align}
&G_{\rm e}(\br,i\omega_n) = -\int_0^\beta d\tau {\rm e}^{i\omega_n \tau}
\la
c^{\vphantom\dagger}_{\br\upa}(\tau) c^{\dagger}_{\br\upa}(0) \ra \\
&= -\int_0^\beta d\tau {\rm e}^{i\omega_n \tau} \frac{1}{N} \sum_\bk
\la f^{\vphantom\dagger}_{\bk\upa} (\tau) f^{\dagger}_{\bk\upa}(0)
\ra \la {\rm e}^{i\theta_{\br}(\tau)} {\rm e}^{-i\theta_{\br}(0)}
\ra.\nonumber \end{align} In another word, on the mean field level, in order
to tunnel in an electron one needs to tunnel in both a spinon and a
charge. Using the Hartree-Fock-Bogoliubov mean field solution of the
spinons and cluster mean field solution of the rotors, we find
$G_e(\br,i\omega_n \to \omega+i\delta)$ at zero temperature: \bea G_{\rm
e}(\br,\omega) &=& \frac{1}{N} \sum_{\bk,m} \left[ u_\bk^2 \frac{|\la m|
{\rm e}^{-i\theta_\br} | 0\ra|^2}
{\omega+i\delta - E_\bk - (\E_m - \E_0)} \right.\nonumber \\
&+& \left. v_\bk^2 \frac{|\la 0| {\rm e}^{-i\theta_\br} | m\ra|^2}
{\omega+i\delta + E_\bk + (\E_m- \E_0)} \right], \eea where $|m\ra$
refers to the $m$-th eigenstates of the rotor cluster with
eigenenergy $\E_m$, with $|0\ra$ being the ground state, $\br$ is a
site belonging to the cluster, $u^2_\bk=(1+\xi_\bk/E_\bk)/2$ and
$v^2_\bk=1-u^2_\bk$ are the usual BCS coherence factors. This leads
to the electronic tunneling density of states (TDOS) $N(\br,\omega) = -
{\pi}^{-1} \Im  G_{\br,\rm e} (\omega)$, \bea N(\br,\omega)
&=& |\Phi|^2 \frac{1}{N}\sum_k [u^2_\bk\delta(\omega-E_\bk)+v^2_\bk
\delta(\omega+E_\bk)] \nonumber\\
&+& \frac{1}{N} \sum_{\bk,m\neq 0} \left[ u_\bk^2 |\la m|{\rm
e}^{-i\theta_\br}|0\ra|^2 \delta(\omega \!\!-\!\! E_\bk\!\! -\!\!
\E_{m0})
\right. \nonumber \\
&+& \left. v_\bk^2 |\la m|{\rm e}^{+i\theta_\br}|0\ra|^2
\delta(\omega\!\! +\!\! E_\bk\!\! +\!\! \E_{m0}) \right].
\label{cluster-N} \eea where $\E_{m0}=\E_m-\E_0$. The first term
proportional to $\Phi^2$, called the coherent part, simply reflects
the Bogoliubov quasiparticle spectrum with energy $E_\bk$. It
describes processes in which the tunneling charge joins the ground state
(the condensate) with no energy cost. This leads to the usual V-shaped
nodal spectrum at low energy characteristic of the d-wave symmetry.
The second term is the incoherent part, corresponding
to processes creating an incoherent charge excitation whose local
energetics is well captured by the cluster mean field theory, and it
dominates at high energies.

\begin{figure}
\includegraphics[width=3.5in]{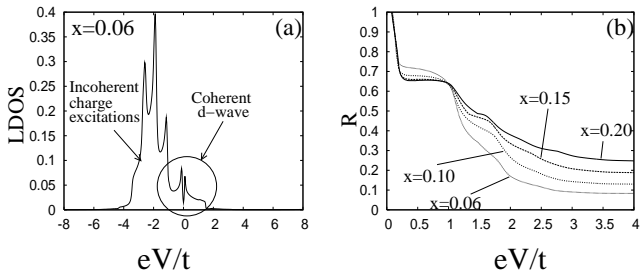}
\caption{(a): The tunneling density of states (TDOS) at doping $x=0.06$.
(b): The dependence of the particle-hole asymmetry ratio $R$ 
(defined in Eq. \ref{eq-R})
on the cutoff energy $\Omega_c$ for a series of dopings.}
\label{fig-tunnel}
\end{figure}

The calculated TDOS is shown in the left panel of Fig. \ref{fig-tunnel}
for $U=12t$, $J=t/3$, and doping $x=0.06$.
Within the cluster mean field approximation, the incoherent weight
consists of discrete peaks corresponding to the excitation energy
levels of the rotor cluster. We have included a small Lorentzian
broadening for these discrete levels. We find that the
tunneling spectrum exhibits a pronounced particle-hole asymmetry:
there are more significant incoherent charge excitations at
negative bias than large positive bias. Roughly speaking, upon extracting 
an electron from
the system, the nearby electrons which are within a few lattice
spacings can lower their kinetic energy,
by $\sim t$, by rattling around the
created empty space. On the contrary, it costs a lot
of energy to add electrons to the system unless one finds an
unoccupied site, and hence one has
little incoherent spectral weight on the positive low bias side.
Qualitatively similar ideas have been proposed by others 
using variational wavefunctions or exact
diagonalization of the $t$-$J$ model on small clusters 
\cite{tunnelingrefs}.

This particle-hole asymmetry can be quantified by introducing the ratio 
\cite{mohit-sumrule} $R(\br)$ of the positive to
the negative spectral weights at a given space point $\br$,
\be
R(\br)=\frac{\int_0^{\Omega_c}N(\br,E)dE}{\int_{-\infty}^0N(\br,E)dE},
\label{eq-R} \ee where the cutoff $\Omega_c \ll U$. 

%%%%%%%%%%%%%%%%%%%%%%%%%%%%%%%%%%%%%%%%%%%%%%%%%%%%%%%%%%%%%%%%%%%%%
%%%%%%%%%%%%%%%%%%%%%%%%%%%%%%%%%%%%%%%%%%%%%%%%%%%%%%%%%%%%%%%%%%%%%

The right panel of Fig. \ref{fig-tunnel} shows the $\Omega_c$
dependence of the ratio
$R$ in the doping range $x=0.06-0.2$. For small $\Omega_c$, it starts off 
close to unity for all dopings, reflecting the nearly particle-hole
symmetric d-wave TDOS at low bias. Very rapidly, however, the
incoherent weight present on the occupied side leads to a
suppression of $R$ which plateaus at a value smaller than unity 
($\sim 0.7$) with
little doping dependence. This is in rough agreement with
the data in Ref.~\onlinecite{davis-Rmap}. At bias $1 \lesssim
eV/t \lesssim 3$ there
is a rapid change in $R$ versus the cutoff, beyond which all the
curves appear to saturate. This is consistent with the bulk of the
incoherent weight on the occupied side being centered at an energy
$\sim$ 2$t$. (The wiggles in $R$ versus $\Omega_c$ in Fig.\ref{fig-tunnel}
are due to the finite size cluster method.)

It has been shown out that for electronic states which obey a Gutzwiller 
projection constraint (two electrons are forbidden from occupying 
the same lattice site), there is a rigorous `partial sum rule' 
\cite{sawatzky-sumrule}
which constrains the spectral weight of occupied and unoccupied states
when there is a separation of scales $t \ll \Omega_c \ll U$. 
In the context of tunneling measurements, this translates
\cite{mohit-sumrule} into the result that the ratio of the integrated 
TDOS at positive bias to that at 
negative bias is $2 \delta/(1-\delta)$ where $\delta$ is the density
of doped holes \cite{mohit-sumrule,pwa-asymmetry}. Furthermore,
the ratio $R(\br)$ has the advantage that the unknown
tunneling matrix elements
that determine the relation between the TDOS and the measured conductance, 
which can depend on the coordinate $\br$, cancel out 
\cite{mohit-sumrule} (assuming they are energy independent).
This `partial sum rule' can then be used to estimate the hole density locally 
from tunneling experiments on doped Mott insulators \cite{mohit-sumrule,
davis-Rmap}. 
The slave boson and rotor mean field theories violate this `partial sum rule' 
since the spinons and rotors are assumed to be independent particles
at mean field level. For $U/t \to \infty$,
the ratio $R$ thus saturates, as a function of $\Omega_c$, to a value 
which is close to $\delta (1+\delta)/(1-\delta)^2$.
This violation is expected to be fixed by including gauge fluctuations 
about the mean field theory \cite{lee-sumrule-fixing}.

\section{Correlated normal states in a weakly doped Mott insulator
on the square lattice}

The finite temperature normal state of underdoped high temperature
superconductors, for $T > T_c$, 
is the least understood part of the phase diagram. 
It exhibits pseudogap phenomenon and highly anomalous thermodynamic
and transport properties. It has been suggested that understanding the
underlying $T=0$ normal state in this regime might shed light on the
pseudogap phenomena.
Very recent experiments by
Doiron-Leyraud {\it et al} \cite{Taillefer-pockets} have begun to
access this ground state in clean crystals of
underdoped oxygen-ordered YBa$_2$Cu$_3$O$_{6.5}$. Applying high
magnetic fields, in the range $\sim 40$-$60$T, to suppress 
superconductivity, they
discovered quantum oscillations in the electrical resistance at
low temperatures. This discovery can be most simply understood by
postulating a well-defined Fermi surface made of small pockets
in the field induced normal ground state. The authors argue that 
the data is less consistent with a specific band structure feature
\cite{band}
of ortho-II YBCO, and more in line with a state with broken
symmetry giving rise to Fermi surface reconstruction.
This state is supposed to compete with superconductivity, to be close 
in energy to the superconducting state, and to show up as the ground
state when superconductivity is strongly suppressed.

Several such candidate states have been proposed. One is the
commensurate SDW at ordering vector $(\pi,\pi)$, which has been
argued to be relevant to the electron doped cuprate superconductors
\cite{millis}.
Given the proximity to the antiferromagnetic Neel phase at low doping, it
is likely that SDW is stabilized and extends to higher doping 
under high fields. This is supported for example by the observation
of antiferromagnetism inside the vortex cores \cite{mitrovic-nmr,core-nmr} 
and from theoretical studies of the
vortex structure using
Bogoliubov-deGennes (BdG) equations \cite{ghosal}. 
Another candidate is
the U(1) staggered flux state \cite{marston-flux} (also known as the DDW
state \cite{ddw-nayak,ddw}).
This state is also well known to be close in energy with the d-wave 
superconducting state \cite{ivanov-lee}
and has been advocated by Patrick Lee and collaborators to describe the 
pseudogap phase \cite{Lee-rmp}.

In
this section we study the SDW state and the U(1) staggered flux state
using cluster slave rotor mean field theory. We then compute the 
expected cyclotron mass $m^*$
for each state as a function of doping, and compare them with
the experiment of Doiron-Leyraud {\it et al} \cite{Taillefer-pockets}. 
We show that by 
studying the doping dependence of the $m^*$ one can, in principle,
unambiguously differentiate between these two scenarios 
for the normal state.
We also suggest a possible resolution for 
the observed difference 
\cite{Taillefer-pockets}
between the hole density as inferred from the
quantum oscillation measurements and from the $T_c$ value.

\subsection{Spin density wave state}

Applying the SDW ansatz Eq. \eqref{ansatz-sdw}, we obtain the mean field spinon
Hamiltonian\bea H_{\rm afm}&=&\sum_{\bk\sigma} \xi_\bk
f^\dagger_{\bk\sigma} f^{\vphantom\dagger}_{\bk\sigma} - 2 J m
\sum_{\bk} \left(f^\dagger_{\bk\upa}
f^{\vphantom\dagger}_{\bk+\bQ\upa}
\right. \nonumber \\
&-& \left. f^\dagger_{\bk\dna} f^{\vphantom\dagger}_{\bk+\bQ\dna}
\right), \eea where we have defined \bea
\xi_\bk &=& -2 (t B + 3 J \chi/4)(\cos k_x + \cos k_y) \nonumber \\
&-& 4  t'B' \cos k_x \cos k_y - \mu_f, \\
\lambda_\bk^{\pm} &=& \frac{1}{2}\left( \xi_\bk + \xi_{\bk+\bQ}
\right)
\pm \frac{1}{2} E_\bk, \\
E_\bk &=& \sqrt{(\xi_\bk - \xi_{\bk+\bQ})^2 + (4 J m)^2}. \eea The
mean field self-consistency conditions then read \bea 1 &=&
\frac{1}{N} \sum_{\bk} \frac{2 J}{E_\bk}
\left[\Theta(-\lambda_\bk^-) - \Theta(-\lambda_\bk^+)\right] ,\\
1-\delta&=&\frac{1}{N} \sum
\left[\Theta(-\lambda_\bk^-) + \Theta(-\lambda_\bk^+)\right] ,\\
\chi&=&\frac{1}{4 N} \sum_\bk \left(\frac{\xi_\bk -
\xi_{\bk+\bQ}}{E_\bk}
\right) \left(\cos k_x + \cos k_y \right) \nonumber \\
&\times& \left[\Theta(-\lambda_\bk^-)-
\Theta(-\lambda_\bk^+)\right]. \eea After self-consistency is
achieved, we can calculate the second neighbor spinon hopping
amplitude using \be \chi' = \frac{1}{2 N} \sum_\bk \cos k_x \cos k_y
\left[\Theta(-\lambda_\bk^-)+ \Theta(-\lambda_\bk^+)\right]. \ee and
use this in solving the rotor model.

In Fig.\ref{fig-normalstate} we plot the sublattice magnetization $M$ as 
function of
doping predicted by the cluster slave rotor mean field theory. $M$ starts at 
the classical Neel state value $1/2$ at half filling and drops monotonically 
with increasing doping and vanishes at $\delta\sim 0.17$. The doped
SDW state has four Fermi pockets centered at the nodal points $\bk_n=(\pm 
\pi/2, \pm \pi/2)$. Naively, from a weak-coupling picture, one expects 
additional electron-like pockets centered around the $(\pi,0)$ point for
small doping. We find that these do not appear in our self-consistent
ground state, suggesting that the weak coupling argument does not work
in this regime. Consistent with this, the area enclosed by the
hole pockets around $(\pi/2,\pi/2)$ satisfies the Luttinger's theorem for
a broken symmetry state with a doubled unit cell.

\subsection{Staggered Flux state}
The U(1) staggered flux ansatz Eq. \eqref{eq-flux-ansatz} leads to the
mean field spinon Hamiltonian
\begin{align}
&H_{\rm flux}=-\sum_{\bk,\sigma}\left[\frac{D_\bk}{2}(f^{\dagger}_{\bk,\sigma}+f^{\dagger}_{\bk+\bQ,\sigma})(f_{\bk,\sigma}-f_{\bk+\bQ,\sigma})+\mathrm{h.c.}\right] \nonumber \\
&-(t'B'\gamma'_\bk+\mu_f)\sum_{\bk,\sigma}
(f^{\dagger}_{\bk,\sigma}f_{\bk,\sigma}+f^{\dagger}_{\bk+\bQ,\sigma}f_{\bk+\bQ,\sigma}),
\end{align}
where $\gamma'_\bk=4\cos k_x \cos k_y$, and the complex hopping \be
D_\bk\equiv
\sum_{\hat{\delta}=\pm\hat{x},\pm\hat{y}}(tB_{i,i+\hat{\delta}}+\frac{3J}{4}\chi_{i+\hat{\delta},i}).
\ee $H_{\rm flux}$ is easily diagonalized to yield the the energy
spectrum, $E^\pm_\bk=-t'B'\gamma'_\bk-\mu_f\pm|D_\bk|$. The ``$+$"
and the ``$-$" branch touches each other at the nodal points
$\bk_n$ regardless the values of staggered
flux, and the low energy excitations are massless Dirac fermions. 
In the doping regime of our interest only the ``$-$"
branch is occupied. The spinon chemical potential $\mu_f$ and the complex 
``order parameter" 
$\chi\equiv \chi_{i,i\pm\hat{x}}=\chi^*_{i,i\pm\hat{y}}$ are found by solving 
the following self-consistency equations \bea
1-\delta=\frac{1}{N}\sum_{\bk}[\Theta(-E^+_\bk)+\Theta(-E^-_\bk)], \\
\chi=\frac{1}{2N}\sum_\bk\cos
k_x\frac{D^*_\bk}{|D_\bk|}[\Theta(-E^-_\bk)+\Theta(-E^+_\bk)]. \eea
Here the sum over $\bk$ is within the reduced Brillouin zone, and
$N$ is the number of sites on each sublattice. Finally, $\chi'$ is obtained
by \be \chi'=\frac{1}{2N}\sum_\bk\cos k_x \cos
k_y[\Theta(-E^+_\bk)+\Theta(-E^-_\bk)]. \ee

%%%%%%%%%%%%%%%%%%%%%%%%%%%%%%%%%%%%%%%%%%%%%%%%%%%%%%%%%%%%%%%%%%%%
\begin{figure}
\includegraphics[width=3.5in]{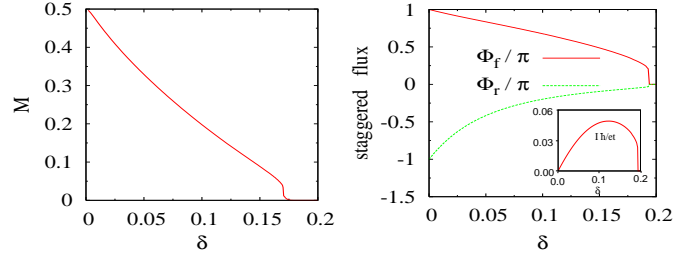}
\caption{Left: The 
sublattice magnetization $M$ in the SDW normal state. 
Right: The spinon and rotor fluxes in the U(1) staggered flux (or DDW) state.
Inset shows the electron current on a nearest neighbor
bond in units of $e t/\hbar$ versus doping.}
\label{fig-normalstate}
\end{figure}
%%%%%%%%%%%%%%%%%%%%%%%%%%%%%%%%%%%%%%%%%%%%%%%%%%%%%%%%%%%%%%%%%%%%%

With self-consistency achieved in the coupled rotor-spinon problem, 
we obtain $\chi=|\chi|e^{\Phi_f/4}$
and $B=|B|e^{\Phi_r/4}$, where $\Phi_f$ is the spinon flux per
plaquette and $\Phi_r$ is the rotor flux (we work in a symmetric
gauge, e.g., $B_{i,i\pm\hat{x}}=B^*_{i,i\pm\hat{y}}$). The orbital
bond current  \be I = \frac{iet}{\hbar} \sum_{\sigma}\langle
c^{\dagger}_{i+x,\sigma}c_{i,\sigma}
-h.c.\rangle=\frac{4et}{\hbar}|\chi
B|\sin(\frac{\Phi_f}{4}+\frac{\Phi_r}{4}) \ee  can be
taken as the order parameter of the U(1) staggered flux (or DDW)
phase. Fig.\ref{fig-normalstate} shows the doping dependence of
$\Phi_f$, $\Phi_r$, and the magnitude of the orbital current. At
half filling the system is in the $\pi$ flux state, $\Phi_f$ and
$\Phi_r$ have the same magnitude $\pi$ but opposite sign,
accordingly the bond current is zero. The current $I$ increases with doping, 
reaching maximum
at $\delta\simeq 0.12$. For a single plaquette, the maximal bond current 
corresponds approximately to a magnetic field of $150$ Gauss at the center 
of the plaquette. $I$ drops to zero around $\delta\sim 0.2$,
where the U(1) staggered flux state gives way to the Fermi liquid state and 
both $\Phi_r$  and $\Phi_f$ vanish.
The total area enclosed by the hole pockets increase linearly
with doping at small doping and is consistent with Luttinger's theorem 
for a broken symmetry state with a doubled unit cell. 

\subsection{Cyclotron mass}
We next discuss whether the quantum oscillation measurements of 
Ref.~\onlinecite{Taillefer-pockets} can shed more light on the 
non-superconducting normal state at low doping, and, in particular,
distinguish between a commensurate SDW metal and the U(1) staggered 
flux (or DDW) state. The quantum
oscillation data offer two pieces of information if the field induced
normal state is a conventional Fermi liquid.
(although more exotic possibilities cannot
be ruled out \cite{rkaul,altman}). 
The first is the area of the hole pocket which is
obtained from the oscillation frequency, and the second is the cyclotron
mass which is obtained from the temperature dependence of the oscillation
amplitude and which is related to the density of states at the Fermi
level. In our calculations, both the SDW and the U(1) staggered flux state have 
hole pocket areas consistent with the Luttinger theorem for a broken symmetry 
state with 
a doubled unit cell and this information cannot be used to discriminate
between them. We therefore turn to a calculation of the cyclotron 
mass in these two states to compare with experiments.

\begin{figure}
\includegraphics[width=3.4in]{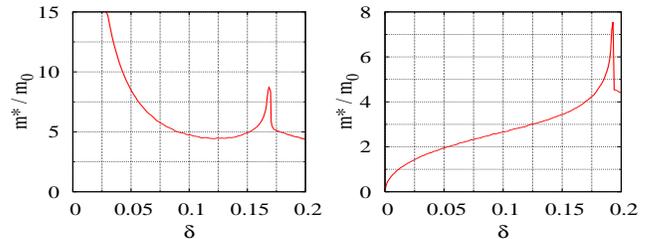}
\caption{Cyclotron mass in units of the bare electron mass in the
SDW state (left) and the U(1) staggered flux (or DDW) state (right).}
\label{fig-cymass}
\end{figure}

The cyclotron effective mass $m^*$ is related to the density of states
at the Fermi level in a Fermi liquid state,
\be
m^* =N_0\frac{2\pi \hbar^2}{ta^2}. \label{def-mstar}
\ee
Here $N_0$ is the density of states per hole pocket per spin at the Fermi 
level measured in units of $1/t a^2$.
As before, we set the hopping $t=350$meV and the Cu-Cu lattice spacing 
$a=3.8\mathrm{\AA}$. The doping dependence of $m^*$ in the SDW state is
shown in the left panel of Fig.\ref{fig-cymass}. The SDW state is 
characterized by a diverging $m^*$ as $\delta\rightarrow 0$. This reflects 
the classical character of the magnetically ordered insulator at
$\delta=0$, namely the staggered magnetization of the insulator in the
mean field theory is $M=0.5$ and the spinon kinetic energy is zero. This
leads to a divergent effective mass for $\delta \to 0$. Spin
fluctuations beyond mean field theory are expected to cut off this 
divergence; nevertheless the
mass is expected to be a strongly increasing function as $\delta$
decreases. The peak in $m^*$ around $\delta\sim 0.17$ is a
feature associated with the vanishing of the SDW order and arises from the
large density of states around the $(\pi,0)$ point.
The magnitude of $m^*$ in the range $\delta=0.1$-$0.15$ 
is around $4.5 m_0$ ($m_0$ being the bare electron mass), 
in apparent disagreement with the experimental value $~1.9 m_0$.

%%%%%%%%%%%%%%%%%%%%%%%%%%%%%%%%%%%%%%%%%%%%%%%%%%%%%%%%%%%%%%%%%%%%%
%%%%%%%%%%%%%%%%%%%%%%%%%%%%%%%%%%%%%%%%%%%%%%%%%%%%%%%%%%%%%%%%%%%%%

The doping evolution of the cyclotron effective mass $m^*$ in the U(1) 
staggered flux (DDW)
state is shown in the right panel of Fig.\ref{fig-cymass}.
This state is characterized by a square root increase of $m^*$, i.e.
$m^*\propto \sqrt{\delta}$, at small doping. This is a direct consequence of 
the linear Dirac spectrum at low energies and therefore a linear density of 
states. Again, the peak feature at $\delta\sim 0.18$ is associated with the 
sudden disappearance of the staggered flux state. 
The value of $m^*\simeq 3 m_0$ at $\delta=0.1$ is closer to the 
experiment result $~1.9 m_0$. 

While a comparison of the cyclotron effective mass with experiments
may seem to favor the staggered flux state over the SDW state as a 
normal state candidate, we cannot rule 
out the SDW scenario based on this quantitative comparison. As we have
seen in previous sections, slave particle 
theory is not quantitatively accurate at the mean field level and
underestimates the nodal $v_F$ in the 
superconducting state by a factor of two at $\delta=0.1$. 
To distinguish these two scenarios, it is more useful to study the
{\it doping dependence} of $m^*$.
From Fig.\ref{fig-cymass} we observe that the effective cyclotron mass 
in the SDW state and the staggered flux state have almost opposite doping 
dependences, we propose that measurements of $m^*$ at a series of
nearby dopings, especially at smaller $\delta$,
may help to identify which candidate normal state is
realized.

Finally, the experiments \cite{Taillefer-pockets} find a difference 
between the carrier densities as inferred from the quantum
oscillation period and from $T_c$.  Assuming 2 hole 
pockets in a reduced Brillouin zone, which is consistent with a SDW or
a U(1) staggered flux state, the oscillation period leads to
an estimate $\delta \sim 0.075$ while the value estimated from
$T_c$ is closer to $\delta \sim 0.1$. This discrepancy
\cite{Taillefer-pockets}, an apparently significant
violation of the Luttinger count, can be accounted for in several
ways including a possible bilayer splitting \cite{nicolas}. However,
the bilayer splitting is expected to be very small at low doping
due to renormalization by strong correlation effects. We offer another
plausible explanation for this
discrepancy. The field induced `normal state' is likely to
contain vortices, which can be inferred from the sign change, with 
temperature,
of the Hall coefficient, a feature which is seen in the mixed state
in other cuprate materials and at other dopings \cite{hagen}. 
It has been suggested that the underlying
normal state of the cuprates should appear in the vicinity of such 
vortices
\cite{vortexcore}. However, it has also been found in BdG calculations 
\cite{ghosal}
of the structure of a single vortex that the density around the
vortex is not uniform. At 
low doping and even in the presence of moderate Coulomb interactions
\cite{ghosal}, 
the density appears to be depleted within the vortex core and piled up as a 
screening cloud at
some distance away from the core, with the far region being superconducting
and at the bulk density.
If the normal state being explored arises from the depleted core region,
the observed density could deviate from the
bulk density in the correct direction. Testing this explanation calls
for further experiments as well as
detailed theoretical calculations of the vortex structure.

\subsection{Implications for Fermi arcs}
We end this section by commenting on the relation between the hole
pockets revealed in the low temperature quantum oscillation experiments
and the Fermi arcs observed in the high temperature normal state
above $T_c$ in photoemission experiments. Assuming that the field
induced normal state has hole pockets, it has been suggested that the
Fermi arcs seen in photoemission studies \cite{arcsexpts} may be just these
hole pockets.
However, the quantum oscillation experiments are carried out at rather 
high magnetic fields $B \sim 50 T$ at very low temperature $T \ll T_c$,
while the photoemission experiments are done in zero magnetic
field and
at fairly high temperatures $T \sim 40-200 K$. Given this, there is no 
{\it a priori}
reason to expect that the photoemission experiments are accessing
the properties of such an underlying normal ground state.
The Fermi arcs seen in the photoemission experiments 
seem to be more consistent with fluctuating d-wave superconductivity
\cite{arcs1,arcs2}. In particular, earlier work by us \cite{arcs1}
suggests an explanation which relies on phenomenological spin charge 
separation ideas, quite close to the route explored in this paper,
and provides a possible explanation for Fermi arcs
whose length scales $~T/T^*$, where $T^*$ is the pseudogap temperature.
Current photoemission experiments do not appear to have the
resolution to settle the issue of whether the arcs indeed arise
from hole pockets. We suggest that further high resolution photoemission
experiments could focus on the momentum distribution of the low
energy spectral weight (integrated from zero to some small energy).
It would be useful to check whether this spectral weight tends to be 
centered at momenta corresponding to 
the low temperature d-wave nodes of the SC, or whether it tends to shift 
away towards some underlying hole pocket Fermi surface once the
system is heated to temperatures above $T_c$.

\section{Discussion}

We have formulated a cluster slave rotor mean field theory
of strongly interacting electronic systems, which extends earlier work
of Florens and Georges \cite{Florens}. In this formulation, the rotor 
sector and spinon sector are coupled to each other by self-consistency
equations, and the rotor Hamiltonian is solved on a finite cluster 
self-consistently coupled to an order parameter bath (rest of the lattice). 
The main advantage of the cluster mean field theory lies in that the short 
range correlation functions of the rotors is properly taken into account. 
By applying it to several examples, we have shown that this method
provides a unified framework to study strongly correlated electronic
systems.
Although we only focused on zero temperature, it is straight forward to 
generalize the theory to finite temperature. Extensions to inhomogeneous
states are also likely to be of interest. 

One shortcoming of the cluster mean field theory for the rotor sector 
is that, by working with finite clusters coupled to a static bath, we 
freeze the Goldstone modes (phase fluctuations). Such Goldstone mode 
fluctuations can be explicitly retained in complementary 
methods such as the random phase approximation \cite{sheshadri} or 
other approaches \cite{huber} to the rotor sector. These methods of solving 
the rotor problem can be combined with the spinon sector to formulate the 
self-consistent slave rotor mean field theory \cite{erhaiunpub}. 

In the path integral formulation of the slave rotor approach 
\cite{Florens,Lee-lee,kim}, our 
mean field theory is equivalent to a saddle point approximation which
ignores the gauge field fluctuations. Such fluctuations
beyond mean field theory are expected to introduce effective interactions 
between slave particles to yield the correct low energy sum rules 
\cite{lee-sumrule-fixing}
including the `partial sum rule' for the particle-hole asymmetry ratio $R$.
These will be explored elsewhere \cite{erhaiunpub}.

%% changed: added note
{\it Note added}: After our paper was submitted, some related works 
have appeared on the arxiv. The self-consistent cluster mean field theory 
for bosons has been shown to successfully describe the supersolid phase 
and the phase diagram of bosons on a triangular lattice \cite{hassan}. 
Two further experiments
carried out on YBa$_2$Cu$_4$O$_8$ (hole doping $\sim 0.15$) have also 
found evidence of Shubnikov-de Haas oscillations \cite{Yelland, Bangura}
in a magnetic field with a larger cyclotron mass, $m^* \sim 2.7-3 m_e$,
than in ortho-II YBCO. A 
recent preprint by Chen et al discusses the 
field induced antiferromagnetic metallic ground state in the underdoped 
cuprates \cite{zhang-rice}. Their principal result is the relation
between the quantum oscillation frequency and the hole doping in this
state and it is in agreement with our conclusion. 
Various proposals for interpreting the Fermi arcs above $T_c$, as well as 
their connection to the hole pockets in ground state, have been examined 
recently by Harrison et al \cite{harrison} and Norman et al \cite{norman-arc}.

\acknowledgments
This research was supported by NSERC and the A.P. Sloan foundation.
We acknowledge stimulating conversations and exchanges with E. Altman, 
Y.-B. Kim, N. Harrison, S. R. Hassan, T. Senthil, and A.-M. Tremblay.  
We are particularly grateful to J. C. Davis, 
N. Doiron-Leyraud, and L. Taillefer for useful discussions about their 
experiments.


\begin{thebibliography}{999}
\bibitem{deficiency}
It is a deficiency of our mean field slave rotor approach that the
value of $J$ has to be seperately specified, and multispin
interactions, which arise naturally within a $t/U$ perturbation
theory in the insulating phase of the Hubbard model, would need to
be included by hand. However, the advantage of the mean field theory
is that we can address the Mott insulator transition while the $t/U$
perturbation is necessarily restricted to the insulating phase.
\end{thebibliography}
\end{document}